\documentclass[conference]{IEEEtran}
\IEEEoverridecommandlockouts
\usepackage{cite}
\usepackage{amsmath,amssymb,amsfonts}
\usepackage{algorithmic}
\usepackage{graphicx}
\usepackage{textcomp}
\usepackage{xcolor}
\def\BibTeX{{\rm B\kern-.05em{\sc i\kern-.025em b}\kern-.08em
    T\kern-.1667em\lower.7ex\hbox{E}\kern-.125emX}}
\begin{document}

\title{Big Data and Analytics Implementation in Tertiary Institutions to Predict Students Performance in
Nigeria\\
}

\author{\IEEEauthorblockN{Ozioma Collins Oguine}
\IEEEauthorblockA{\textit{Department of Computer Science} \\
\textit{University of Abuja}\\
Abuja, Nigeria \\
oziomaoguine007@gmail.com}
\and
\IEEEauthorblockN{Kanyifeechukwu Jane Oguine}
\IEEEauthorblockA{\textit{Department of Computer Science} \\
\textit{University of Abuja}\\
Abuja, Nigeria\\
jeaniettee@gmail.com}
\and
\IEEEauthorblockN{Hashim Ibrahim Bisallah}
\IEEEauthorblockA{\textit{Department of Computer Science} \\
\textit{University of Abuja}\\
Abuja, Nigeria\\
hashim.bisallah@uniabuja.edu.ng}
}

\maketitle

\begin{abstract}
The term, ‘Big Data’ has been coined to refer to the gargantuan bulk of data that cannot be dealt with by traditional data-handling techniques. Big Data is still a novel concept, and in the following literature we intend to elaborate it in a palpable fashion. It commences with the concept of the subject in itself along with its properties and the two general approaches of dealing with it. Big
Data provides an opportunity to educational Institutions to use their Information Technology resources strategically to improve educational quality and guide students to higher rates of completion, and to improve student persistence and outcomes. This paper explores the attributes of big data that are relevant to educational institutions, investigates the factors influencing adoption of big data and analytics in learning institutions and seeks to establish the limiting factors hindering use of big data in
Institutions of higher learning. A survey research design was adopted in conducting this research and Questionnaire was the instrument employed for data collection.
\end{abstract}

\begin{IEEEkeywords}
BigData, Data Analytics, Education, Data Science, AI in Education, KDD, Machine Learning
\end{IEEEkeywords}

\section{Introduction}
Imagine a world without data storage; a place where details about people or organizations, transaction performed, and every aspect which can be documented is lost directly after use. Organizations would thus lose the ability to extract valuable information and knowledge, perform detailed analyses, as well as provide new opportunities and advantages. Anything ranging from customer names and addresses, to products available, to purchases made, to
employees hired, has become essential for day-to- day continuity. Data is the building block upon which any organization thrives.

Now think of the extent of details and the surge of data and information provided nowadays through the advancements in technologies and the internet. With the increase in storage
capabilities and methods of data collection, huge amounts of data have become easily available. Every second, more and more data
are being created and needs to be stored and analyzed in order to extract value. Furthermore, data has become cheaper to store, so
organizations need to get as much value as possible from the huge amounts of stored data.

Data is characterized as the lifeblood of decision-making and the raw material for accountability. Without high-quality data
providing the right information on the right things at the right time, designing, monitoring and evaluating effective policies becomes
almost impossible (United Nations, 2014). In that context, an ongoing attention to data and data-driven approaches from academics and professionals exists, since the knowledge arising from data analysis processes leads to the promotion of innovative activity, transforming organizations, enterprises and national economies.

Nowadays, in the 4th Industrial revolution era, organizations and governments focus on the development of capabilities that provide
knowledge extracted from large and complex data sets, commonly known as “big data”. Big data is a buzzword in the last years in the business and economics fields, since it plays an essential role in economic activity and has strengthened its role in creating economic value by enabling new ways to spur innovation and productivity growth. Hence, the ability of management, analysis and acting is significant under the context of knowledge-based capital (KBC) that is associated with digital information, innovative capacity and economic aspects (OECD, 2015).

In that era, many enterprises independent size, from start-ups to large organizations, attempt to obtain data-driven culture struggling for competitive advantage against rivals. Enterprises aim to leverage data generated within organizations through their operations to gain valuable insights for better, faster and more accurate decisions in crucial business issues.

The tremendous generation of data, expected to reach 180 ZB in 2025, give data a leading role in change and growth of the 21st-century
shaping a new “digital universe” with the transformation of markets and businesses (Economist T., 2017). Digital information from complex and heterogeneous data coming from anywhere and at any time introducing a new era, the era of “Big Data” (Sivarajah, U. et al,
2017).

Manyika, J., et al. (2011) viewed Big data as large datasets that are not able to be captured, stored, managed and analyzed by typical
software tools. These data sets that are huge - not only in size- but also in heterogeneity and complexity (structured, semi-structured and unstructured data) including operational, transactional, sales, marketing and other data. In addition, big data includes data that comes in several formats including text, sound, video, image and more. This unstructured data is growing faster than structured and have captured the $90\%$ of all the data (Gantz, J., Reinsel, D., 2011). Therefore, new forms of processing capabilities are required for getting data insights that lead to better decision making.

\section{STATEMENT OF PROBLEM}
As with other IT-related initiatives, Big Data also has its own set of problems and challenges. The Economic Intelligence Unit research
mentioned earlier indicates some of the impediments to the effective utilization of Big Data for decision-making (Capgemini, 2012).
McAfee and Brynjolfsson (2012) indicate five management challenges which prevent organizations from reaping the full benefits of Big Data utilization. They are: leadership, talent management, technology, decision making, company culture. When considering leadership, having more or better data does not guarantee success. The leaders still have to have a vision of the organization’s development, set clear goals, understand the market, etc. Big Data changes the way organizations make many of their decisions.\\
Talent management is connected with the necessity of providing the organization with the right people (such as data scientists) who are
prepared to work with huge sets of data. The next challenge relates to the problem of assuring the data scientists have the proper tools
to handle the Big Data. Although the technology alone is not enough to succeed in Big Data initiatives, it is a necessary part of it.
The next challenge is connected with the problem of ensuring there are mechanisms in place to guarantee that the information and the
relevant decision-makers are in the same location. It is important to make sure that the people who understand the problems are able to
use the right data and to work with people who have the necessary problem-solving skills.\\

The final challenge is connected with changes related to organizational culture. The key issue in this context is to make decisions as data-driven as possible, instead of basing them on
hunches and instinct (E. Brynjolfsson, A. McAfee, 2012). It is worth mentioning that the significance of such cultural transformation is
also mentioned in other research for instance in the area concerning sectoral Big Data projects.

In addition, the numerous challenges connected with data and legal rights should be noted. They relate to such issues as copyright, database rights, confidentiality, trademarks, contract law, competition law (Kemp Little, 2013). There is another important challenge also connected with legal aspects. It relates to the transparency in data collection practices. A further important risk is around the utilization of Big Data to increase the automation of decision-making. \\

There is one more important danger which is underlined in the context of Big Data. It is connected with the fact that Big Data might not be providing the whole picture for a particular situation. There are several reasons for this i.e. biases in data collection, exclusions or gaps in data signals or the constant need for context in conclusions (Ferguson, 2013).

\section{SIGNIFICANCE OF THE STUDY}
Big Data is a knowledge system that is already changing the objects of knowledge and social theory in many fields while also having the
potential to transform management decision-making theory (Boyd \& Crawford, 2012). Big Data incorporates the emergent research field of
learning analytics (Long \& Siemen, 2011), which is already a growing area in education.

However, research in learning analytics has largely been limited to examining indicators of individual student and class performance. Big
Data brings new opportunities and challenges for institutions of higher education. Long and Siemen (2011) indicated that Big Data presents the most dramatic framework in efficiently utilizing the vast array of data and ultimately shaping the future of higher education. The application of Big Data in tertiary institutions
was also echoed by Wagner and Ice (2012), who noted that technological developments have certainly served as catalysts for the move towards the growth of analytics in tertiary institutions of higher education.

In the context of higher education, Big Data connotes the interpretation of a wide range of administrative and operational data gathered processes aimed at assessing institutional performance and progress in order to predict future performance and identify potential issues related to academic programming, research, teaching and learning (Hrabowski, Suess \& Fritz, 2011a, 2011b; Picciano, 2012). Others indicated that to meet the demands of improved
productivity, higher education has to bring the tool of analytics into the system. As an emerging field within education, a number of
scholars have contended that Big Data framework is well positioned to address some of the key challenges currently facing higher
education (Long \& Siemens, 2011).

\section{LITERTURE REVIEW}

Big data has been around us since much longer than we think, and in order to better understanding the potential of this great new
jump in technology and analytics it is essential to understand the challenging problems that motivated the now so called "data scientists" to conceive and develop solutions and technologies that now drive this revolution. Data in the 21st Century is like Oil in the 18th Century: it is an immensely, untapped valuable asset. Like oil, for those who see Data’s fundamental value and learn to extract and use it there will be huge rewards.

Big data and analytics (BDA) continue to spark interest among scholars and practitioners. Organizations are increasingly aware that they may process and analyze their large data volumes to capture value for their businesses and employees (George, Haas and Pentland, 2014). With the advent of more computational power, machine learning – particularly deep learning through neural networks – has become more broadly deployable in organizations.
Academic research on the topic also skyrocketed. When searching for the term “big data”, the Web of Science Core Collection yields 3,347 hits in 2015, and over 4,000 in both 2016 and 2017.//

In its most basic form, Big data and Analytics is the “extensive use of data, statistical and quantitative analysis, explanatory and
predictive models, and fact-based management to drive decisions and actions” (Davenport \& Harris, 2007). Big Data and Analytics involves
the integration, extraction, and transformation of data into actionable information so that specific decision points are identified. “Action” is an important element of the definition. Data
analytics is more than analysis. It is a form of communication in which the analysis is transformed into a recommended action to
guide decision-making. Within the field of data analytics, it is not enough to simply produce data and reports.

Scholars have argued that novel machine learning capabilities may realize the predictive value of big data, unleashing its strategic
potential to transform business processes and providing the organizational capabilities to tackle key business challenges (Fosso Wamba et al., 2015). Yet, very few attempts have been made to culminate the plethora of BDA research and explore the underlying theoretical foundations. Although some attempts have been made to review and theorize how organizational value can be derived from BDA, these attempts have mostly taken on a rather narrow information systems and technology perspective (for some exceptions Günther et al., 2015). Calls to explore the organizational impact of BDA from other functional management perspectives remain largely unanswered to this date. \\

Petersen (2012) connoted that an important component of data analytics is a focus on process. Data analytics has the greatest
potential when it is developed and embedded in normalized business processes. Liberatore \& Luo (2011) is of the view that a well-developed analytics program translates data into analysis, analysis into insight, and insight into managerial actions, such as improving
operational decisions, redesigning or changing existing processes, formulating or adjusting strategies, or improving decision quality and speed. According to Vassakis (2018), Big data can be
characterized by the seven V's: volume, variety, veracity, velocity, variability, visualization and value.
\paragraph{\textbf{Volume:} this refers to the large size of the datasets. It is fact that Internet of Things (IoT) through the development and increase of connected smartphones, sensors and other devices, in combination with the rapidly developing Information and Communication
Technologies (ICTs) including Artificial Intelligence (AI) have contributed to the tremendous generation of data (counting records, transactions, tables, files etc.). The speed of data is surpassing Moore’s law and the volume of data generation introduced new
measures for data storage i.e. exabytes, zettabytes and yottabytes.}

\paragraph{\textbf{Variety:}  this represents the increasing diversity of data generation sources and data formats. Web 3.0 leads to growth of web and social media networks leading to the generation of different types of data. From messages, updates, photos and videos that are posted in social media networks like Facebook or Twitter, SMS, GPS signals from smartphones, customers transactions
in banking, e-business and retail, voice data in call centers etc. Many of the crucial sources of big data are comparatively novel, including mobile devices that supply huge streams of data that are connected with human behavior through their activities and locations; or web sources supplying data through comprising logs, clickstreams and social media actions. Additionally, big data also differs in data types that are generated, thus big data consists on structured data (tables, records), unstructured data (text and voice), semi-structured data (XML, RSS feeds) and other data that is difficult to classify like data deriving from audio, video and other appliances.}

\paragraph{\textbf{Variability:}  this is often confused with variety, but variability is related with rapid change of meaning. For instance, words in a text can have a different meaning according to context of a text, thus for an accurate sentiment analysis, algorithms need to find out the meaning (sentiment) of a word taking into account the whole context.}

\paragraph{\textbf{Velocity:}  Velocity: Big data is characterized by the high speed of data generation. Data generated by
connected devices and web arriving in enterprises in real-time. This speed is extremely significant for enterprises in taking various actions that enable them to be more agile, gaining competitive
advantage against competitors. Despite the fact that some enterprises have already exploited big data (click-streams data) to offer their customers purchase recommendations, nowadays enterprises
though big data analytics have the ability to analyze and understand data taking actions in real-time.}

\paragraph{\textbf{Veracity:} Veracity of data refers to data
reliability and accuracy. The data collection has data that are not clean and accurate, thus data veracity refers to the data uncertainty and the level of reliability correlated with some type of
data.
} 

\paragraph{\textbf{Visualization:}  Data visualization is the science
of visual representation of data and information. It presents quantitative and qualitative information in some schematic form, indicating patterns, trends, anomalies, constancy, variation, in ways that cannot be presented in other forms like text and tables (Gantz \& Reinsel, 2011). The leverage of big data can provide valuable knowledge and thus the value offered by the data analysis process can benefit enterprises, organizations, communities and consumers.
Enterprises that overcome challenges and exploit big data efficiently have more precise information and are able to create new
knowledge by which they can improve their strategy and business operations regarding well defined targets like productivity, financial performance and market value (Friendly, 2008), while big data plays a major role in digital transformation of enterprises introducing innovations. Therefore, an increasing interest in exploitation of big data among enterprises and organizations exist.
}
\section{RELEVANCE / SIGNIFICANCE AND OPPORTUNITIES OF BIG DATA AND
ANALYTICS IMPLEMENTATION IN TERTIARY INSTITUTIONS}
Big Data in tertiary institutions of higher education also covers database systems that store large quantities of longitudinal data on
students’ right down to very specific transactions and activities on learning and teaching. When students interact with learning
technologies, they leave behind data trails that can reveal their sentiments, social connections, intentions and goals. Researchers can use such data to examine patterns of student performance
over time that is from one semester to another or from one year to another.

From an organizational learning perspective, it is well understood that institutional effectiveness and adaptation to change relies on
the analysis of appropriate data (Rowley, 1998) and that today’s technologies enable institutions to gain insights from data with previously unachievable levels of sophistication, speed and accuracy (Jacqueline, 2012). As technologies continue to penetrate all facets of higher education, valuable information is being generated by students, computer applications and systems (Hrabowski \& Suess, 2010).
\begin{figure}[htbp]
\centerline{\includegraphics[width=\columnwidth,]{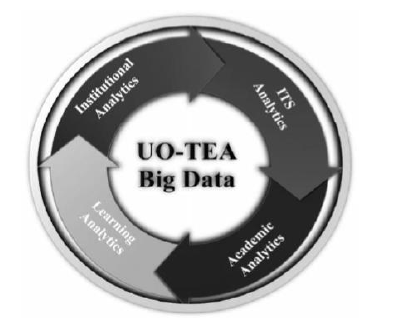}}
\caption{Conceptual framework of a typical University Analytics
Technology}
\label{4}
\end{figure}

Furthermore, Big Data Analytics could be applied to examine student entry on a course assessment, discussion board entries, blog entries or wiki activity, which could generate thousands of transactions per student per course. These data would be collected in real or near real time as it is transacted and then analyzed to suggest courses of action. As Siemens (2011) indicated that “[learning] analytics are a foundational tool for informed change in education” and provide evidence on which to form understanding and make informed (rather than instinctive) decisions.

The most important benefits for Big Data in tertiary education are as follows:

\paragraph{Improved Instruction:} Can improve students’ performance and learning abilities making the lessons more personal. The courses
can be improved from the teachers with the help of analytics.
\paragraph{Matching Students to Programs:} Big Data is able to help students to find the best educational program.
\paragraph{Matching Students to Employment:} Companies and candidate employees can discover alternative and more effective tools to
use big data to qualify their skills with the needed skills. In addition, students can find and make applications for jobs that can match with their abilities, more efficient than before.
\paragraph{Transparent Education Financing:} Students can participate in education activities, which previously they don’t have the ability. Furthermore, being able to choose anything
about higher education and to discover the most proper education programs for them.
\paragraph{Efficient System Administration:} Education systems are able to develop a skills education supply that can help administrators to allow more effective educational resources. In
that way, this secures a high performance and afford to a versatile and smart plan for future education interests.

\section{CHALLENGES OF BIG DATA AND ANALYTICS IMPLEMENTATION IN
TERTIARY INSTITUTIONS }
Furthermore, data integration challenges are eminent, especially where data come in both structured and unstructured formats and needed to be integrated from disparate sources most of
which are stored in systems managed by different departments. Additionally, data cleansing when performing integration of
structured and unstructured data is likely to result to loss of data.

There are also challenges associated with quality of data collected and reported. Lack of standardized measures and indicators make
inter(national) comparison difficult, as the quality of information generated from Big Data is totally dependent on the quality of data
collected and the robustness of the measures or indicators used. administrative data, classroom and online data can pose additional challenges (Daniel \& Butson, 2013).

\section{RESEARCH METHODOLOGY} 
Data is a cumulative resource which isn’t just abundant but effectively infinite and durable. The need to collect, analyze, provide answers to questions, test hypothesis, make sound judgement and accurate decisions gave rise to data analysis.
Data analysis entails inspecting, cleaning, transforming and modelling data with the goal of discovering useful information, informing conclusion and supporting decision making. It plays a major role in decision making and helping businesses operate more effectively.\\

This section is aimed at analyzing the impact of the internet to student of tertiary institution through data analysis approach. The data analysis phase of this research work was executed using Jupyter Notebook. The Jupyter Notebook is an open source application and a
python framework that supports a wide range of workflows in scientific programming, numerical simulation, statistical modelling, data visualization, machine learning and so much more. It allows you to create and share document that contains codes visualization,
equations, and comments. The programming language on which this analysis phase was carried out was Python Programming. Python is
a general-purpose, interactive, interpreted, easy to use and object-oriented scripting language. Python has a host of robust standard libraries which gives users the option of choosing from wide range of modules based on the specific need such as panda, numpy, matplotlib, etc. \\

This section of this research is aimed at: Generating a data source of predictive variables, Identifying the different factors affecting students’ performance in Tertiary Institutions Constructing a prediction model using classification data mining techniques on
the basis of identified predictive variables and Validating the developed model for tertiary institution students.

\section{DATA ANALYSIS PROCESSES}
There are various steps to data mining and analysis. The steps employed in this research work is outlined below.

\subsection{Data Extraction:}  
The relevant dataset used in this research for prediction was collected through questionnaire distributed digitally to different students within their groups chat as an online survey using
Google Forms, the data was collected anonymously and without any predisposition.
The initial size of the dataset is 205 records. The attributes used were: age, gender, department, ethnicity, academic records and
lesson duration. The dataset used were prepared using Microsoft Excel, saved with a .csv file extension in order to be read.

\subsection{Data Preparation for Modelling
(Tokenization):}  This involves the process of gathering,
combining, organizing and structuring data to be analyzed. data preparation comprises of those techniques concerned with analyzing data in its raw form in order to yield quality data.
These techniques include data collection, profiling, data cleaning, data validation, data transformation and data. During this process,
data are pulled from different internal systems and external sources. The key purpose of data preparation is to ensure that datasets to be used for analysis is accurate and consistent. Data
preparation generates datasets smaller than the original one which will significantly improve the efficiency of data analysis. It generates high quality data, which leads to quality pattern.

\subsection{Data Visualization:}  Data visualization also known as data exploration is one of the steps in data analysis. It entails representing information and data graphically using visual and statistical elements like charts, maps, graphics, plots, graphs and
other tools. Data visualization tools provides a better way to see and understand complex trends, correlations and patterns in dataset that might go undetected in text-based forms and also to communicate information clearly, effectively and efficiently. In order to
understand the dataset in hand, it must be explored in a statistical manner, as well as, visualize it using graphical plots and diagrams.
This step-in data mining is essential because it allows the researchers as well as the readers to understand the data before jumping into applying more complex data mining tasks and
algorithms.

\begin{figure}[htbp]
\centerline{\includegraphics[width=\columnwidth,]{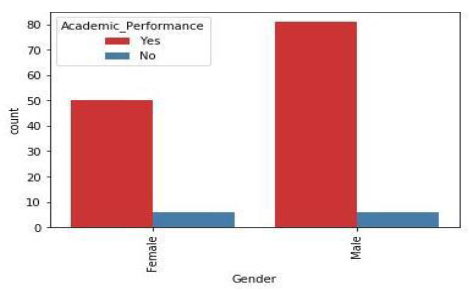}}
\caption{Model Visualization based on Gender}
\label{2}
\end{figure}

\begin{figure}[htbp]
\centerline{\includegraphics[width=\columnwidth,]{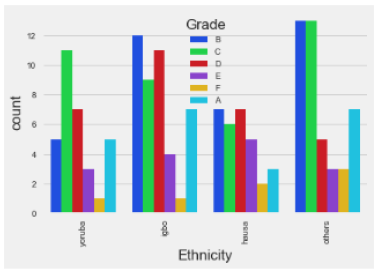}}
\caption{Model Visualization based on Ethnicity}
\label{3}
\end{figure}

\subsection{Creation and Evaluation of Models:}
Classification is a commonly used and studied data mining technique. It is a mostly used and studied classification because it is simple and easy to use. In detail, in data mining, Classification is a technique for predicting a data object’s class or category based on
previously learned classes from a training dataset, where the classes of the objects are known. The classification techniques used in training the employs the use of mathematical algorithms in evaluating datasets; It includes: Decision Trees, Ada Boost, Support Vector Regression, Naïve Bayes and Stochastic gradient Descent.

The selection of the algorithms was based on the best and most used algorithms for analysis and prediction. Models were evaluated using
the cross-validation method with stratified sampling.

Four classification models (Decision Trees, Ada Boost, Support Vector Regression, Naïve Bayes and Stochastic gradient Descent) were
trained and tested. Using the cross-validation score as a basis of evaluation it can be deduced that two models have the same cross validation and occupies the highest cross validation amongst all four trained and tested model which are the Naïve Bayes classification model and the Ada Boost Model.

Therefore, either the Naïve Bayes or Ada Boost approach will be adopted and best suited for the analysis.

\begin{figure}[htbp]
\centerline{\includegraphics[width=\columnwidth,]{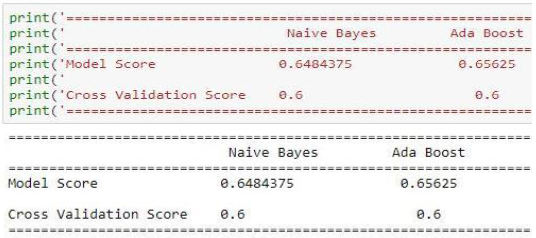}}
\caption{Model Comparison after Training}
\label{4}
\end{figure}

\section{INTERPRETATION FROM ANALYSIS}
The strengths points are good enough to start. There are many sources of data in the tertiary institutions, which can be funding the Big Data to analyze and investigate them. The basic infrastructure exists including automated systems and databases systems. The social media is there as well and will be a vital source
to add value to the analysis. On the other hand, the weakness points can be grouped into two sections, financial weakness and cultural
impact. \\

Most tertiary institutions still suffer from some financial problems due to the country’s economic problem. This problem affects the
establishment of the infrastructure, the training programs, and the ability to hire qualified employees. In addition, culture still affects the Big Data and Analytics implementation. The resistance of sharing information from the students and universities' academic staff is still considered one of the biggest problems regarding Big Data. When we come to the opportunities that will be provided by Big Data and Analytics, we see many chances to improve education and develop the decision-making process and change the way the learning is conduct. The various types of analytics can bring light to the quality of educations. \\

However, the threats, which may affect the implementation of Big Data and Analytics, can be overcome by putting the project under the
responsibilities of the professionals. Build and create the environment that will encourage all students, academic staff from all colleges to participate and make use for all the multimedia,
will be the most way to support Big Data and Analytics.

\section{SUMMARY}
The rapidly growing amount of data which organizations have at their disposal and the opportunities connected with its practical
utilization are increasingly changing the processes relating to making decisions at various organizational levels. Thus, Big Data
offers huge potential to positively impact on the functioning of tertiary Institutions generally and gives them a competitive advantage. Most institutes of higher education are now trying to
utilize to an even greater degree the opportunities and chances that are emerging. \\
But if initiatives aimed at the practical usage of Big Data and Analytics are to be successful at giving tertiary institutions a competitive advantage and be of value, it is not enough to just collect and own the appropriate data sets. In fact, this is only the starting point of every Big Data initiative. Further essential elements are suitable analytical models, tools, skilled people,
and organizational capabilities. Lack of all of these necessary components can lead to a situation whereby instead of expected benefits there is only disappointment and a belief that
Big Data initiatives are only the next wave in a long line of management fads.

Generally, although Big Data solutions have a huge potential for both commercial organizations and governments, there is
uncertainty concerning the speed with which they can be utilized in a secure and useful way.

\section{CONCLUSION}
In conclusion, Big Data does not bring new data; it just makes use of what is there and analyses explore and presents the knowledge.
This advanced analysis from Big Data and Analytics brings many benefits to the students, academic staff, and administration staff. A
SWOT analysis was established to find the strengths and weakness of Tertiary Institutions with Baze University, Abuja as a case study.
Opportunities and threats from the adoption of Big Data and Analytics was also examined.
\section{DISCUSSION AND FUTURE WORK}
Overall, this research investigated that Big data and Analytics plays a great role in the tertiary education system. \\
Much work, however, needs to be done in terms of refining the process applied to the data. In particular, the log data should be time-sliced and processed on a per period basis in order to determine whether and how the student’s level and type of activity changes over time. An extended approach for further enriching the
VSM-based activity vector was also proposed by processing the datetime and IP address metadata of the log data. This enriched vector representation can be used as input to any classification/predictive model. An eventual application of this model is the immediate identification of at-risk students based on the actions they are exhibiting in the online environment. \\

Experiments testing the enriched representation on several machine learning algorithms using Python and the Scikit-learning library were also performed. The results indicate that classification algorithms can modestly predict a student’s academic performance and, in particular, model the difference between high,
low, and failed performances. This modest result indicates that there are more factors that need to be considered in predicting the
performance of students in blended courses.

More powerful machine learning classification techniques can be tested on the enriched vector representation to further isolate these factors and to determine whether the classification
accuracy can still be improved.

\section{Disclosure}
The authors declare that they have no competing interests.

\section{Author Contributions}
All authors contributed significantly to conception and design, data acquisition, analysis, and interpretation; participated in compiling the article and critically revising it for significant intellectual content; agreed to the submission.

\vspace{12pt}

\end{document}